\documentclass[10pt,conference,final,twocolumn]{IEEEtran}
\usepackage[T1]{fontenc}
\usepackage{cite}
\usepackage{amsthm}
\usepackage{amssymb}
\usepackage{amsfonts}
\usepackage{fancyhdr}  %
\usepackage{extarrows}
\usepackage{algorithm}
\usepackage{multirow,tabularx}
\usepackage{mathtools}
\usepackage{xcolor}
\usepackage[english]{babel}
\usepackage{caption}
\usepackage{bm}
\usepackage{algorithmic}
\usepackage{amsmath}
\usepackage{subfigure}
\usepackage{makecell}
\usepackage{colortbl}
\usepackage{booktabs}
\usepackage{amsmath}
\usepackage{graphicx}
\usepackage{diagbox}
\usepackage{cancel}
\usepackage{tablefootnote}
\usepackage{makecell}

\usepackage{hyperref} 
\usepackage{cleveref} 
\usepackage{setspace}
\hypersetup{colorlinks = true, 
	    linkcolor = black, 
	    urlcolor = black,
            citecolor = black} 
\begin{document}
\bstctlcite{IEEEexample:BSTcontrol}  
\title{
Synergizing Global Pattern Learning and Time Order Characterization in Mobile Channel Prediction: An RWKV-Based Approach
}
\author{
Zili~Wang\textsuperscript{$\dagger$},
Zirui~Chen\textsuperscript{$\dagger$},
Ridong~Li\textsuperscript{$\dagger$},
Zhaoyang~Zhang\textsuperscript{$\dagger$},
and Zhaohui~Yang\textsuperscript{$\dagger$}\\
\IEEEauthorblockA{
    \textsuperscript{$\dagger$}College of Information Science and Electronic Engineering, Zhejiang University, Hangzhou 310027, China\\
    \textsuperscript{$\dagger$}Zhejiang Provincial Laboratory of Multi-Modal Communication Networks \\and Intelligent Information Processing, Hangzhou 310027, China\\
    E-mails: \{zili\_wang, ziruichen, lrd, ning\_ming, yang\_zhaohui\}@zju.edu.cn
}
}

\maketitle
\begin{abstract}
Owing to the potential to reduce pilot overhead and mitigate channel aging, channel prediction is emerging as an important research topic in wireless communications. Meanwhile, deep neural networks are becoming a foundational technology for high-precision prediction thanks to their excellent non-linear representation capabilities. 
In this paper, we conceive a task-driven prediction network, which aims to deeply synergize the following two functions: learning global patterns for shareable features across adjacent time slots and structurally encoding time order to characterize the inherent causality within the channel dynamics. 
To implement channel prediction accuracy, we employ RWKV (receptance weighted key value) as network backbone and adapt it to the task's specific characteristics, utilizing its deep interleaved learning architecture to extract global patterns across multiple channel samples and leveraging its unique exponential decay to characterize temporal order.
These task-driven unique designs significantly improve the learning efficiency of prediction network.
Comprehensive experimental evaluations demonstrate the superiority of the proposed method over current data-driven methods, such as long short-term memory and Transformer, in the channel prediction task, including 1.84\textasciitilde4.29 dB gains in normalized mean squared error and 2.6\textasciitilde10.5 percentage point gains in cosine correlation.
\end{abstract} 
\begin{IEEEkeywords}
  Channel prediction, MIMO-OFDM, deep learning, neural network, RWKV.
\end{IEEEkeywords}

\IEEEpeerreviewmaketitle

\section{Introduction}\label{section1}
In wireless communication systems, timely acquiring accurate channel state information (CSI) is important for the formulation of appropriate configuration parameters. By predicting current CSI based on past channel samples, channel prediction approach significantly alleviates the excessive pilot overhead and channel aging problems, thus becoming a research topic of great value \cite{CD}.

Traditional signal processing-based methods, such as parametric \cite{parametric} and autoregressive models \cite{selfregretion}, mathematically model wireless channel with interpretability. However, their reliance on simplified assumptions renders these methods ill-suited for non-linear and highly dynamic real-world communication environments. Recently, deep learning (DL) has emerged as an approach attracting significant interest for its improved performance. 
The long short-term memory (LSTM) network in \cite{lstm_cp} employs gating mechanisms to learn data dependencies, demonstrating the efficacy of non-linear DL methods for prediction. Besides, Transformer \cite{transformer_cp}, via its self-attention mechanism, models a stronger global sequence dependency than LSTM. 


Nevertheless, these current researches typically apply advanced network architectures from a data-driven standpoint, overlooking the task's inherent characteristics. In this paper, we start from task-driven perspective, arguing that prediction network performance significantly benefits from the deep coupling of two core capabilities: the effective extraction of global patterns via dedicated temporal dependency modules to fully learn multi-scale channel features and the structural temporal encoding to capture the causality underlying the channel's dynamics. However, existing methods fall short in synergizing these two capabilities, as shown in Table \ref{nn_comparison}. Specifically, while the serial recurrent structure of LSTM \cite{lstm_cp} inherently incorporates temporal information, its compression of the entire history into a fixed-size hidden state impairs its ability to capture long-range dependencies. Conversely, the Transformer's \cite{transformer_cp} one-time additive positional encoding at the input layer offers only a shallow temporal characterization. 

\newcolumntype{C}[1]{>{\centering\arraybackslash}m{#1}}
\begin{table}[htbp]\footnotesize
    \centering
    \caption{General comparisons of inductive biases across models} 
    \label{tab:model_comparison_grid}
\renewcommand{\arraystretch}{1.25} 
    \begin{tabular}{l|C{1.5cm}|C{1.5cm}|C{1.5cm}}
        \hline 
        \diagbox{\makecell[l]{Inherent\\property}}{\makecell[l]{Learning\\model}} & 
        LSTM & Transformer & RWKV \\
        \hline 
        \multirow{2}{*}{\makecell[l]{Effective long-range\\dependencies}} & 
        \multirow{2}{*}{{\color{black}\normalsize$\times$}} & 
        \multirow{2}{*}{{\color{black}\normalsize$\checkmark$}} & 
        \multirow{2}{*}{{\color{black}\normalsize$\checkmark$}} \\ & & & \\
        \hline 
        \multirow{2}{*}{\makecell[l]{Structural positional\\encoding}} & 
        \multirow{2}{*}{{\color{black}\normalsize$\checkmark$}} & 
        \multirow{2}{*}{{\color{black}\normalsize$\times$}} & 
        \multirow{2}{*}{{\color{black}\normalsize$\checkmark$}} \\ & & & \\
        \hline 
    \end{tabular}
    \label{nn_comparison}
\end{table}

Accordingly, we introduce receptance weighted key value (RWKV) \cite{rwkv}, an emerging advanced architecture, to the channel prediction task, with necessary adaptations. This architecture is particularly well-suited to the functional requirement of deeply synergizing global pattern learning and structural time-order characterization. The main contributions of this paper are as follows:

\begin{itemize}
    \item We analyze the challenges of learning multi-scale channel features in prediction task and accordingly propose a novel network framework synergizing global pattern learning and structural temporal encoding.

    \item We employ the RWKV model to meet these task-driven requirements, leveraging its deep interleaved architecture to extract global patterns and time-decay mechanism to characterize the causality within channel dynamics.

    \item Experimental evaluations demonstrate the model's competitive performance, particularly in low-data regimes, along with its enhanced robustness and adaptability against noise and variations in data correlations.
\end{itemize}

\section{System Model}\label{section2}
In this section, we present a multiple-input multiple-output orthogonal frequency division multiplexing (MIMO-OFDM) channel model \cite{MIMO} and introduce the problem formulation and physical process of channel prediction task.

\subsection{Channel Model}\label{section2.1}
Considered a MIMO-OFDM system, a base station (BS) with an $N_{\rm{t}}$-antenna uniform linear array (ULA) communicates with multiple single-antenna users. Each channel comprises $P$ propagation paths, whose combined effect determines the channel response on each of the $N_{\rm{c}}$ subcarriers, which is given by:
\begin{equation}
\mathbf{h}\left( f \right) =\sum_{p=1}^P{\alpha _p}e^{-j2\pi f\tau _p}\mathbf{a}\left( \theta _p \right),
\label{eq:hf}
\end{equation}
where $\alpha _p$ is amplitude attenuation, $\tau _p$ is transmission delay, $\theta _p$ is angle of arrival (AoA) of the $p$-th path, and $f$ is the carrier frequency. The specific expression for $\mathbf{a}\left( \theta _p \right)$ is as follows:
\begin{equation}
\mathbf{a}\left( \theta _p \right) =\left[ 1,e^{-j2\pi f\frac{d\cos \theta _p}{c}},\cdots ,e^{-j2\pi f\frac{d\left( N_{\rm{t}}-1 \right) \cos \theta _p}{c}} \right] ^\mathsf{T},
\end{equation}
where $c$ is the speed of light and $d$ is the spacing between adjacent antennas. The transmission distance offset between the $(i+1)$-th antenna and the first antenna is denoted by $di\cos \theta _p$, resulting in a phase offset of $2\pi f\frac{di\cos \theta _p}{c}$. Thus, CSI matrix $\mathbf{H}\in \mathbb{C}^{N_{\rm{t}}\times N_{\rm{c}}}$ between the BS and a single user can be expressed as:
\begin{equation}
\mathbf{H}=\left[ \mathbf{h}\left( f_1 \right) , \mathbf{h}\left( f_2 \right) ,\cdots ,\mathbf{h}\left( f_{N_{\rm{c}}} \right) \right],
\end{equation}
where $f_i=f_1+(i-1)\Delta f, \quad \left( i=1, 2, \cdots, N_{\rm{c}} \right)$, $f_i$ and $f_1$ denote the frequency of the $i$-th and 1st subcarrier respectively, and $\Delta f$ is the frequency spacing between two adjacent subcarriers.


\subsection{Problem Formulation}\label{section2.2}
Assume that the BS has stored the CSI of the past $n$ time steps, denoted by $\left\{ \mathbf{H}_{t-n},...,\mathbf{H}_{t-1} \right\}$. The predicted CSI for the current time step $t$ is denoted by $\mathbf{\hat{H}}_t$. Therefore, the channel prediction problem can be formulated as:
\begin{equation}
\mathbf{\hat{H}}_t=g_{\text{CP}}\left( \mathbf{H}_{t-n},...,\mathbf{H}_{t-1} \right),
\label{eq:CP}
\end{equation}
where $g_{CP}(\cdot)$ represents the channel prediction function.

Channel dynamics is a continuous physical process, governed by smooth variation of the parameters such as $\alpha_p$, $\tau_p$, and $\theta_p$ during user movement. As indicated by Eq.~\eqref{eq:CP}, the historical sequence of channel matrices $\mathbf{H}_{t-n},...,\mathbf{H}_{t-1}$, derived from discrete observations of continuous process, must be fed into the model sequentially, which demonstrates that channel prediction is a strictly time-series task.

Additionally, channel state dynamics is also a multi-scale physical process. Its small-scale fast fading is dictated by the phase term $e^{-j2\pi f\tau _p}$ in Eq.~\eqref{eq:hf}, while its large-scale slow fading is determined by amplitude term $\alpha _p$. Within a localized spatial region, the amplitude $\alpha _p$ remains relatively stable due to high similarity in the scattering environment and multipath structure. In contrast, even subtle changes in propagation delay $\tau _p$ induced by slight user movement are significantly amplified by the large carrier frequency $f$, resulting in rapid phase fluctuations. This implies that small-scale components reflect precise, local channel dynamics, while large-scale components reflect the long-term trends under a broader communication environment.

\section{Proposed Method and Related Analysis}\label{section3}
This section details the proposed RWKV-based channel prediction method, including its task-driven functional requirements, the underlying principles of RWKV, and the complete architectural design.

\begin{figure*}[htbp]
    \centering
    \includegraphics[width=0.94\linewidth, trim=5cm 29cm 25cm 4cm, clip=true]{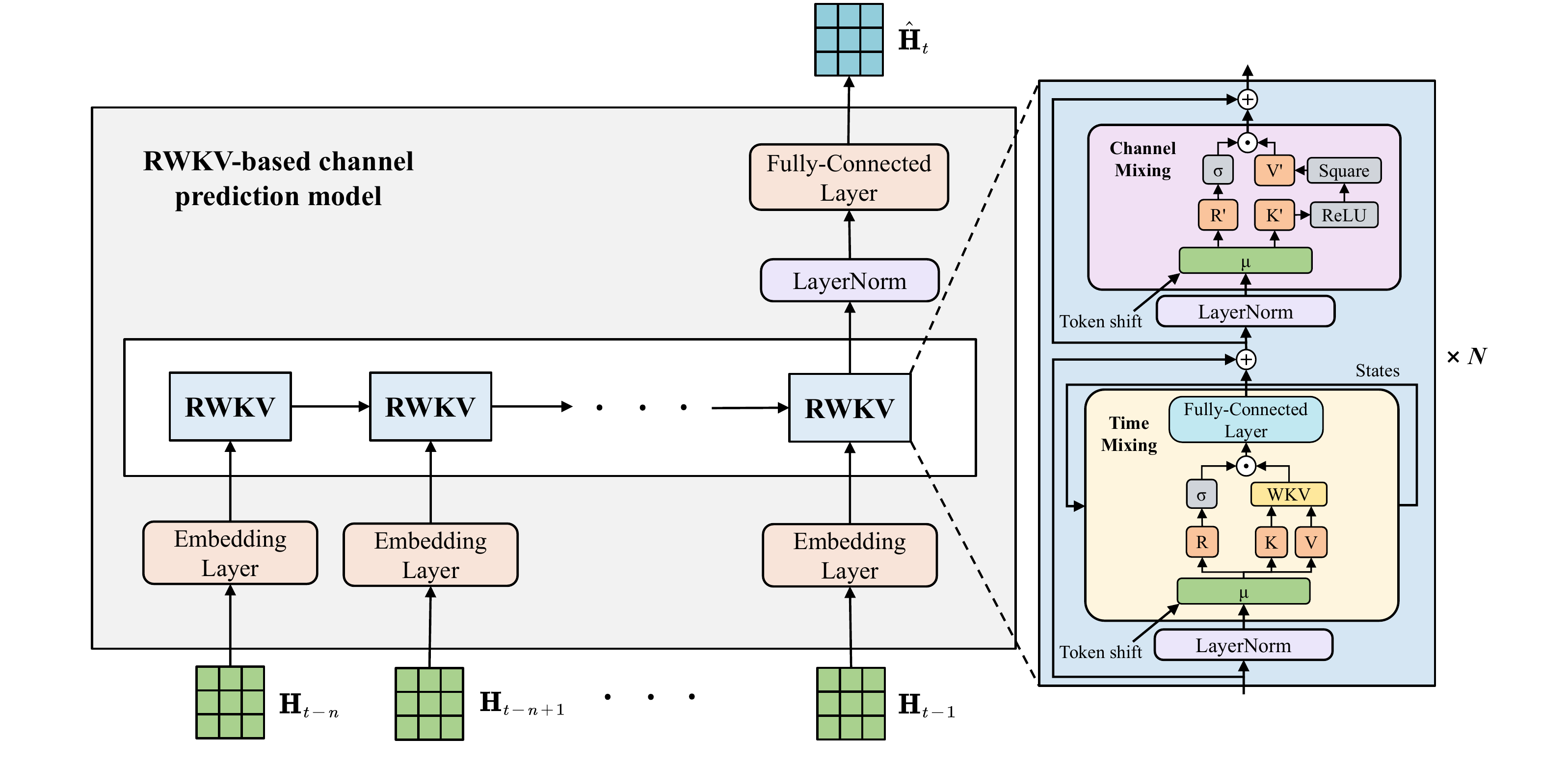}
    \vspace{0.3em}
    \caption{The proposed learning structure of RWKV-based mobile channel prediction.}
    \label{fig.1:RWKV based CP}
    \vspace{-0.5em}
\end{figure*}

\subsection{Requirements of Advanced Prediction Model}\label{section3.1}

Based on the physical characteristics of channel prediction explored in Section~\ref{section2.2}, we now outline the functional requirements for the model to better fit this task. Firstly, by processing input channel matrices $\mathbf{H}_{t-n},...,\mathbf{H}_{t-1}$ sequentially, the model is enabled to learn their step-by-step transitional relationships. This makes it crucial for the model to learn the input sequence order. Although LSTM has a recurrent structure naturally suited for temporal perception, it simultaneously obscures the contribution of specific inputs to current output and thus limits the capture of global dependencies. This limitation motivates the use of explicit positional encoding like that in Transformer. Whereas, applying a single, additive positional encoding only at the input layer is a superficial treatment. Therefore, we require combining structural features with explicit positional encoding to structurally characterize the time-order. Specifically, this should be realized by repeatedly applying a formulaic encoding within each layer to embed positional processing at the more fundamental levels of the model. Meanwhile, beyond merely recognizing sequence order, we require that the positional encoding itself characterize the causal regularities governing channel dynamics.

Moreover, the channel's complete dynamics are composed of large-scale features and small-scale features. To acquire multi-scale features, global pattern learning is employed. By observing the complete history of inputs, which covers a larger spatial range, this approach enables the model to learn multi-scale features more comprehensively, thereby achieving a deeper understanding of the communication environment.
Beyond simply using all historical data, the model must manage the input sequence efficiently. Specifically, the model should be able to assign unique weights to each input based on its content and temporal position. This allows it to adaptively focus on the most critical historical information. Such dynamic allocation scheme enables efficient memory management and strengthens the model's ability to capture long-range dependencies. 

\subsection{Architecture and Characteristics of RWKV}\label{section3.2}
As a novel learning architecture, RWKV \cite{rwkv} has shown promising results in the domain of large language models (LLMs).
Adapted from the attention free Transformer (AFT) \cite{aft}, the time mixing module of RWKV modifies the positional encoding to drastically reduce the parameter count, as defined by the following equations:
\begin{equation}
\boldsymbol{wkv}_t = \frac{\sum_{i=1}^{t-1} e^{-(t-1-i)\boldsymbol{w}+\boldsymbol{k}_i} \odot \boldsymbol{v}_i + e^{\boldsymbol{u}+\boldsymbol{k}_t} \odot \boldsymbol{v}_t}{\sum_{i=1}^{t-1} e^{-(t-1-i)\boldsymbol{w}+\boldsymbol{k}_i} + e^{\boldsymbol{u}+\boldsymbol{k}_t}},
\label{eq:wkv}
\end{equation}

\begin{equation}
\boldsymbol{o}_t=W_o\cdot \left( \sigma \left( \boldsymbol{r}_t \right) \odot \boldsymbol{wkv}_t \right),
\end{equation}
where $\boldsymbol{k}_t$, $\boldsymbol{v}_t$, and $\boldsymbol{r}_t$ are the parameters that can be dynamically computed from the input at time $t$ and $t-1$. The vectors $\boldsymbol{w}$ and $\boldsymbol{u}$ are learnable parameters. Notably, $\boldsymbol{w}$ is a time-decay weight applied to past tokens, while $\boldsymbol{u}$ serves as a direct bonus for the current token. $\odot$ is the element-wise product, $\sigma \left( \cdot \right)$ is the non-linear sigmoid function, and $W_o\left( \cdot \right)$ represents applying an fully-connected (FC) layer. Also, Eq.~\eqref{eq:wkv} can be expressed in an iterative form to achieve serial inference as follows,
\begin{equation}
\boldsymbol{wkv}_t=\frac{\boldsymbol{a}_{t-1}+e^{\boldsymbol{u}+\boldsymbol{k}_t}\odot \boldsymbol{v}_t}{\boldsymbol{b}_{t-1}+e^{\boldsymbol{u}+\boldsymbol{k}_t}},
\label{eq:wkv iterative}
\end{equation}
where $\boldsymbol{a}_t=e^{-\boldsymbol{w}}\odot \boldsymbol{a}_{t-1}+e^{\boldsymbol{k}_t}\odot \boldsymbol{v}_t$, $\boldsymbol{b}_t=e^{-\boldsymbol{w}}\odot \boldsymbol{b}_{t-1}+e^{\boldsymbol{k}_t}$. 

The core of the time mixing module is the time-decay mechanism, as detailed in Eq.~\eqref{eq:wkv}. The mechanism assigns each historical input a unique weight that decays exponentially as its temporal separation from the current step grows. Such a scheme provides both dynamic attention over the global input and an efficient method for history compression, essential for capturing global dependencies. 

This layer-wise application of weights serves to embed time-order awareness in a structural form. Moreover, the design of decaying weights incorporates a physically intuitive prior: the relevance of historical CSI to the current state diminishes over time. The reason is that temporally adjacent CSI instances usually correspond to neighboring areas, leading to similar scattering environments and path structures, and thus higher correlation.


\subsection{Adapting RWKV to Channel Prediction Task}\label{section3.3}
Recognizing the strong alignment between the architectural characteristics of RWKV and the functional requirements for channel prediction, we introduce this model to the prediction task with necessary adaptations.

Standard DL networks typically handle real numbers, whereas channel matrices are complex-valued, with their real and imaginary parts jointly constituting the channel's structural features across different dimensions. Considering that processing real and imaginary parts separately compromises the channel's feature structure \cite{cmixer, FL_IndoorPos}, we stack them as two input dimensions. This allows the network to thoroughly mix information between both parts, enabling the extraction of comprehensive channel features.

Given that RWKV is originally designed for LLMs, we adapt its input and output layers for the channel prediction task. We remove the original tokenizer, word embedding layer, and LM Head, replacing them with a pair of symmetric FC layers that perform inverse operations. This design leverages channel sparsity to map the data to a low-dimensional space, reducing computational complexity. The input embedding layer projects the input channel matrices from $2 \times N_{\rm{t}} \times N_{\rm{c}}$ to the hiddensize dimension, while the output FC layer maps this representation back to the original dimension. As depicted in Fig.~\ref{fig.1:RWKV based CP}, the RWKV-based mobile channel prediction network is then composed of $N$ RWKV blocks and a Layer Normalization (LN) layer.
Each RWKV block contains two sub-modules: time mixing and channel mixing. The time mixing module, as the core of RWKV's temporal processing, governs the information exchange between different time steps. The channel mixing module operates within each individual time step to fuse information across the feature dimensions. This module's function is analogous to that of the feed-forward network (FFN) \cite{transformer} in Transformer, serving to enhance the model's expressive power via non-linear transformation. This process is implemented by the following equations:
\begin{equation}
\boldsymbol{o}_t'=\sigma \left( \boldsymbol{r}_t' \right) \odot \left( W_v'\cdot \max \left( \boldsymbol{k}_t',0 \right) ^2 \right).
\end{equation}
The two sub-modules are interconnected via residual connections and LN, which facilitates gradient propagation in deep networks and stabilizes the training process. Furthermore, states are propagated across time steps to maintain a continuous flow of information. The token shift mechanism, which computes the $\boldsymbol{k}_t$, $\boldsymbol{v}_t$, and $\boldsymbol{r}_t$ vectors by fusing inputs from time steps $t$ and $t-1$, provides the computational prerequisite for both sub-modules. 

The calculation process in the block is as follow: In the Time Mixing module, the $n\times d$-size input obtained from the embedding layer, first undergoes token shift to generate $\boldsymbol{k}_t$, $\boldsymbol{v}_t$, and $\boldsymbol{r}_t$, with computational complexity of $\mathcal{O}(nd^2)$ (where $n$ is the sequence length and $d$ is the hidden size). Then, since $\boldsymbol{wkv}_t$ can be expressed in the iterative form of Eq.~\eqref{eq:wkv iterative}, it incurs only $\mathcal{O}(nd)$ cost. The final output projection via a FC layer adds another $\mathcal{O}(nd^2)$. Thus, the overall complexity of the Time Mixing module is $\mathcal{O}(nd^2)$. The Channel Mixing module begins with token shift similarly ($\mathcal{O}(nd^2)$), followed by element-wise activation ($\mathcal{O}(nd)$). The subsequent value projection and output gating contribute $\mathcal{O}(nd^2)$ and $\mathcal{O}(nd)$, respectively, yielding a module total cost of $\mathcal{O}(nd^2)$.Summing these components, the total computational complexity of $N$ blocks is $\mathcal{O}(Nnd^2)$.
During training, the loss function is computed solely on the current time step's output $\mathbf{\hat{H}}_t$. We use the mean squared error (MSE) as our loss function, given by:
\begin{equation}
\text{MSE\ }=\ \frac{1}{num}\sum_{i=1}^{num}{\lVert \mathbf{H}_i-\mathbf{\hat{H}}_i \rVert _{F}^{2}},
\end{equation}
where $num$ is the number of training samples, and $\lVert \cdot \rVert _{F}$ is the Frobenius norm of a matrix, which is similar to the Euclidean norm of a vector.

\section{Experiments}\label{section4}
In this section, we present the experiment settings and evaluate the performance of our proposed model.
 
\subsection{Experiment Settings}\label{section4.1}
\begin{figure}[htbp]
    \centering 
    \includegraphics[width=0.45\textwidth, trim=4cm 22cm 49cm 15cm, clip=true]{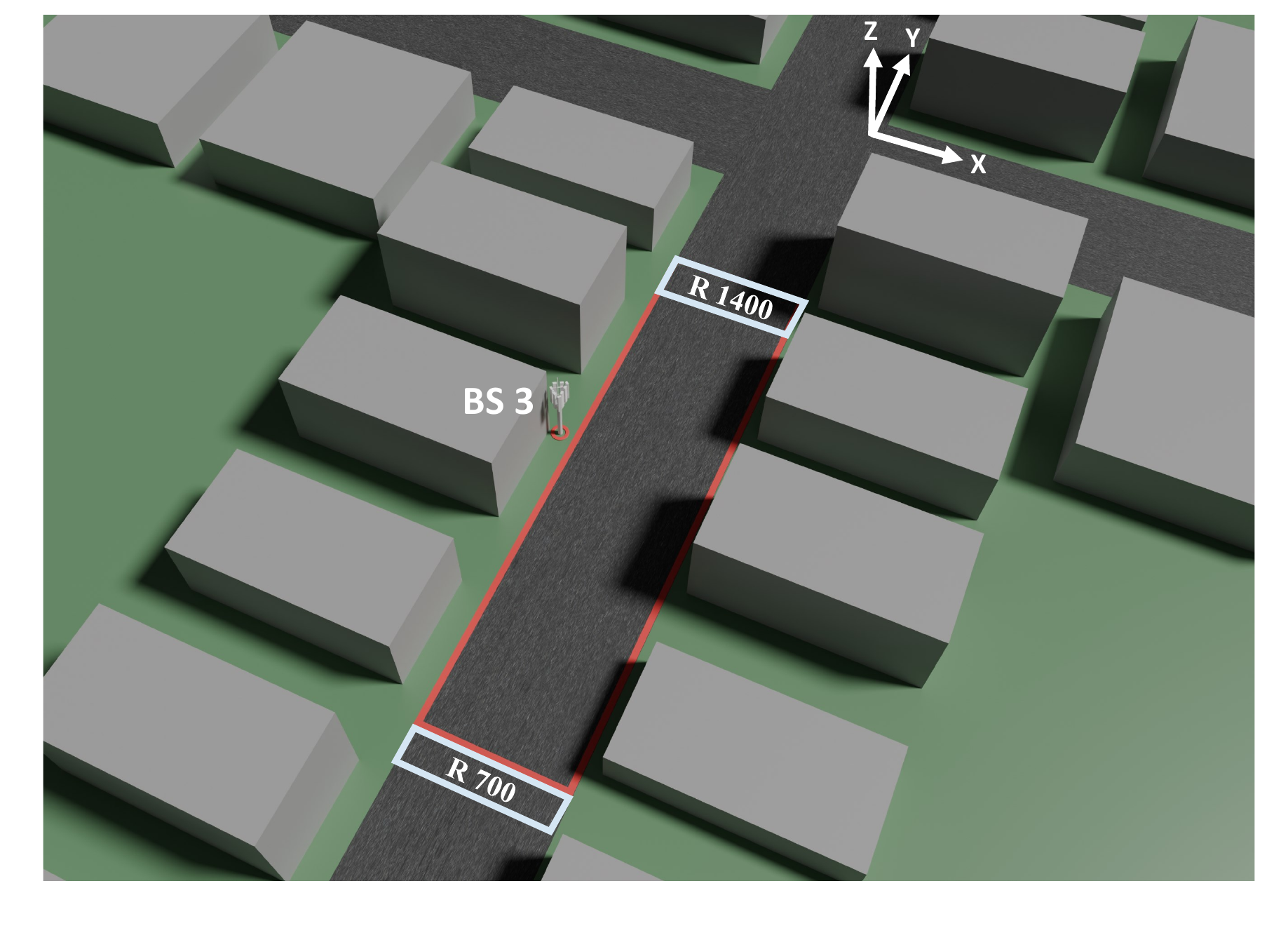} 
    \caption{`O1' scenario in DeepMIMO dataset \cite{deepmimo}.}
    \label{fig.2:O1}
\end{figure}

Our simulation experiments are conducted using the DeepMIMO dataset \cite{deepmimo}, with the detailed parameters of our chosen `O1' urban outdoor scenario presented in Table~\ref{tab:deepmimo_params}. In this setup, BS3 serves users from `R701' to `R1400' within the red-boxed region, as shown in Fig.~\ref{fig.2:O1}.
Since the DeepMIMO dataset provides static channel data, we substitute temporal intervals with spatial intervals to construct the required time-series prediction. We assume that users move at a constant velocity along the $x$-axis, $y$-axis, or a diagonal path. Under this assumption, the CSI of spatially equidistant users is modeled as a uniformly sampled time series. 

To validate the effectiveness of our proposed method, we selected two mainstream channel prediction methods, LSTM\cite{lstm} and Transformer\cite{transformer}, to serve as baselines. Regarding the model scale, we ensured that the parameter count for each baseline model was sufficient to reach its performance saturation, thereby guaranteeing a fair comparison. The specific parameters for all models are listed in Table~\ref{tab:nn_params}. 

\begin{table}[htbp] \footnotesize
  \centering 
  \caption{Parameters setting for DeepMIMO datasets} 
  \label{tab:deepmimo_params} 
  \vspace{0.2em}  
  \begin{tabular}{p{3.5cm} p{4.5cm}}
    \toprule 
    \textbf{Parameters} & \textbf{Value} \\
    \midrule 
    Scenario & O1 \\
    Frequency band & 3.5GHz \\
    Bandwidth & 40MHz \\
    Base station & BS3 \\
    Antenna array form & ULA \\
    Number of antennas ($N_{\text{t}}$) & 32 \\
    Number of subcarriers ($N_{\text{c}}$) & 32 \\
    Number of paths ($P$) & 25 \\
    User area & R701 - R1400 \\
    Area size & 79.8\,m $\times$ 36\,m \\
    Number of training data & \{10000, 20000, 40000, 80000\} (default 40000) \\
    Number of testing data & 20000 \\
    Sequence length  & 8 \\
    \bottomrule 
  \end{tabular}
\end{table}

\begin{table}[htbp]\footnotesize
    \centering
        \caption{Parameters setting of RWKV, Transformer and LSTM}
        \label{tab:nn_params}
        \vspace{0.2em}
        \begin{tabular}{p{3cm} p{5cm}}
            \toprule
            \textbf{Parameters} & \textbf{Value} \\
            \midrule
            Batch size & 400 \\
            Hidden size & 512 \\
            Epochs & 1000 \\
            Loss function & MSE \\
            Optimizer & AdamW \\
            \midrule
            \makecell[l]{Number of blocks} & \makecell[l]{RWKV: 6; \\ Transformer: 5; \\ LSTM: 3} \\
            \midrule
            Learning rate & \makecell[l]{RWKV, Transformer: \\ Linear warmup $\rightarrow$ Cosine decay $\rightarrow$ Constant \\ (warmup\_steps = 15000, cosine\_decay\_steps \\ = 166500, constant\_steps = 81500, \\start\_lr = $1 \times 10^{-3}$, max\_lr = $1 \times 10^{-2}$, \\constant\_lr = $5 \times 10^{-7}$); \\ LSTM: $1 \times 10^{-3}$ (multiply by 0.2 at epochs \\150, 350, 450)}   \\ 
            \bottomrule
        \end{tabular}  
\end{table}

We use the normalized MSE (NMSE) and cosine correlation $\rho$ as the performance indexes, which are formulated as follows:
\begin{equation}
\text{NMSE}=\mathbb{E}\left\{ \frac{\lVert \mathbf{H}-\mathbf{\hat{H}} \rVert _{F}^{2}}{\lVert \mathbf{H} \rVert _{F}^{2}} \right\},
\end{equation}
and
\begin{equation}
\rho =\mathbb{E}\left\{ \frac{1}{N_{\text{c}}}\sum_{m=1}^{N_{\text{c}}}{\frac{|\mathbf{\hat{h}}_{m}^{\text{H}}\mathbf{h}_m|}{\lVert \mathbf{\hat{h}}_m \rVert _2\lVert \mathbf{h}_m \rVert _2}} \right\},
\end{equation}
where $\mathbf{H}$ and $\mathbf{h}_m$ are the ground-truth channel matrix and $m$-th subcarrier vector, while $\mathbf{\hat{H}}$ and $\mathbf{\hat{h}}_{m}$ are their predicted counterparts. And $\lVert \cdot \rVert _{F}$ is the Frobenius norm.

\subsection{Performance Evaluation}\label{section4.2}
\subsubsection{Prediction Accuracy}
To evaluate the prediction accuracy and sample efficiency of our proposed method, we compared our RWKV model against LSTM and Transformer on training sets across four training sets of varying sizes, as summarized in Table~\ref{tab:nmse_rho_accuracy}. The experiment results show that while all models improve with more training data—with the Transformer's performance gains being particularly significant—our proposed RWKV model exhibits leading performance across all training data scales, especially in low-data regime. With 40,000 training samples (this setting maintained for all subsequent experiments), RWKV achieves gains of 1.84\textasciitilde4.29 dB in NMSE and 2.6\textasciitilde10.5 percentage points in $\rho$.
This advantage stems from its stronger inductive bias compared to Transformer, which is attributed to its physically intuitive time-decay mechanism.

\begin{figure}[htbp]
    \centering 
    \includegraphics[width=0.45\textwidth, trim=0.2cm 0.19cm 0.8cm 1.8cm, clip=true]{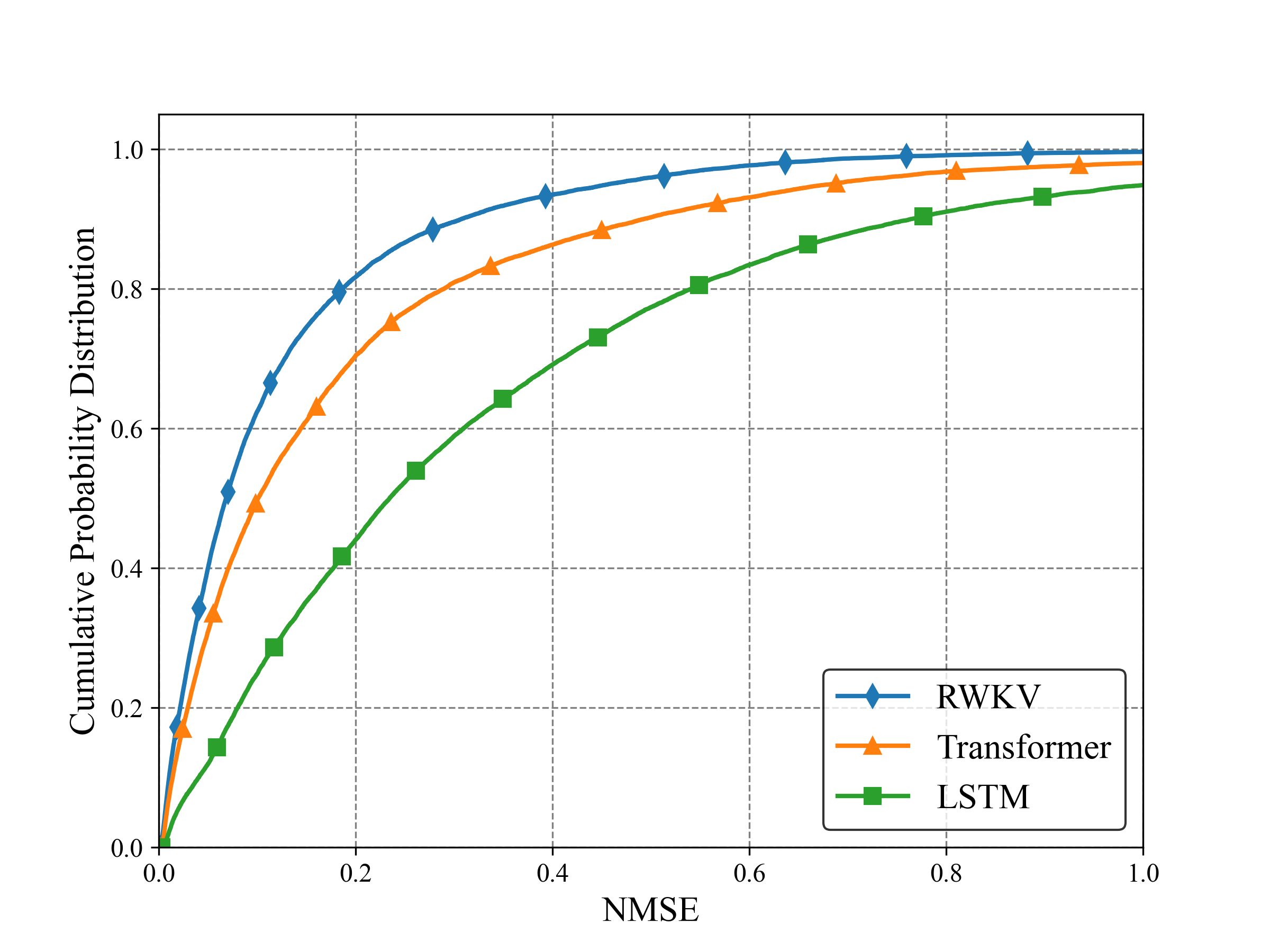} 
    \caption{CDF of NMSE between predicted channel and true channel.}
    \label{fig.3:cdf}
\end{figure}

In addition to average accuracy, we assess the model's generalization across each test sample by calculating the cumulative distribution function (CDF) of the NMSE, presented in Fig.~\ref{fig.3:cdf}. 
Our proposed RWKV model possesses the smallest proportion of high-error samples among the methods. Specifically, approximately 80\% of its prediction errors fall below an NMSE of 0.2, indicating high accuracy and stability.
    
\begin{table}[htbp]
  \centering
  \caption{NMSE and $\rho$ under various training data size.}
  \label{tab:nmse_rho_accuracy}
  \small 
  \vspace{0.5mm}  
  \renewcommand{\arraystretch}{1.3} 
  \begin{tabular}{c|c|c|c|c|c}
    \hline
    \multirow{2}{*}{\textbf{Indexes}} & \multirow{2}{*}{\textbf{Schemes}} & \multicolumn{4}{c}{\textbf{Number of training samples}} \\
    \cline{3-6} 
    & & 10,000 & 20,000 & 40,000 & 80,000 \\
    \hline
    
    \multirow{3}{*}{\shortstack{NMSE \\ (dB)}} & RWKV & \textbf{-3.16} & \textbf{-6.68} & \textbf{-9.13} & \textbf{-10.81} \\
    \cline{2-6}
    & Transformer & -1.22 & -4.22 & -7.29 & -9.73 \\
    \cline{2-6}
    & LSTM & -1.07 & -2.87 & -4.84 & -7.11 \\
    \hline
    
    \multirow{3}{*}{$\rho$} & RWKV & \textbf{0.8745} & \textbf{0.9345} & \textbf{0.9536} & \textbf{0.9626} \\
    \cline{2-6}
    & Transformer & 0.7992 & 0.8804 & 0.9295 & 0.9518 \\
    \cline{2-6}
    & LSTM & 0.6645 & 0.7825 & 0.8629 & 0.9180 \\
    \hline
  \end{tabular}
\end{table}

\subsubsection{Performance under Various Sampling Intervals}
In channel prediction, model predicting relies on learning dependencies within the input data. In this part, we explore how alterations in data correlation affect model performance by changing the sampling interval. We train and compare the proposed RWKV and baselines under short (maintained for all other experiments), medium, and long sampling intervals. Performance consistency on testing samples is quantified by error bars representing the 10th-90th percentile NMSE range, where shorter bars indicate higher stability.

\begin{figure}[htbp]
    \centering 
    \includegraphics[width=0.45\textwidth, trim=0.2cm 0.8cm 0.5cm 1.8cm, clip=true]{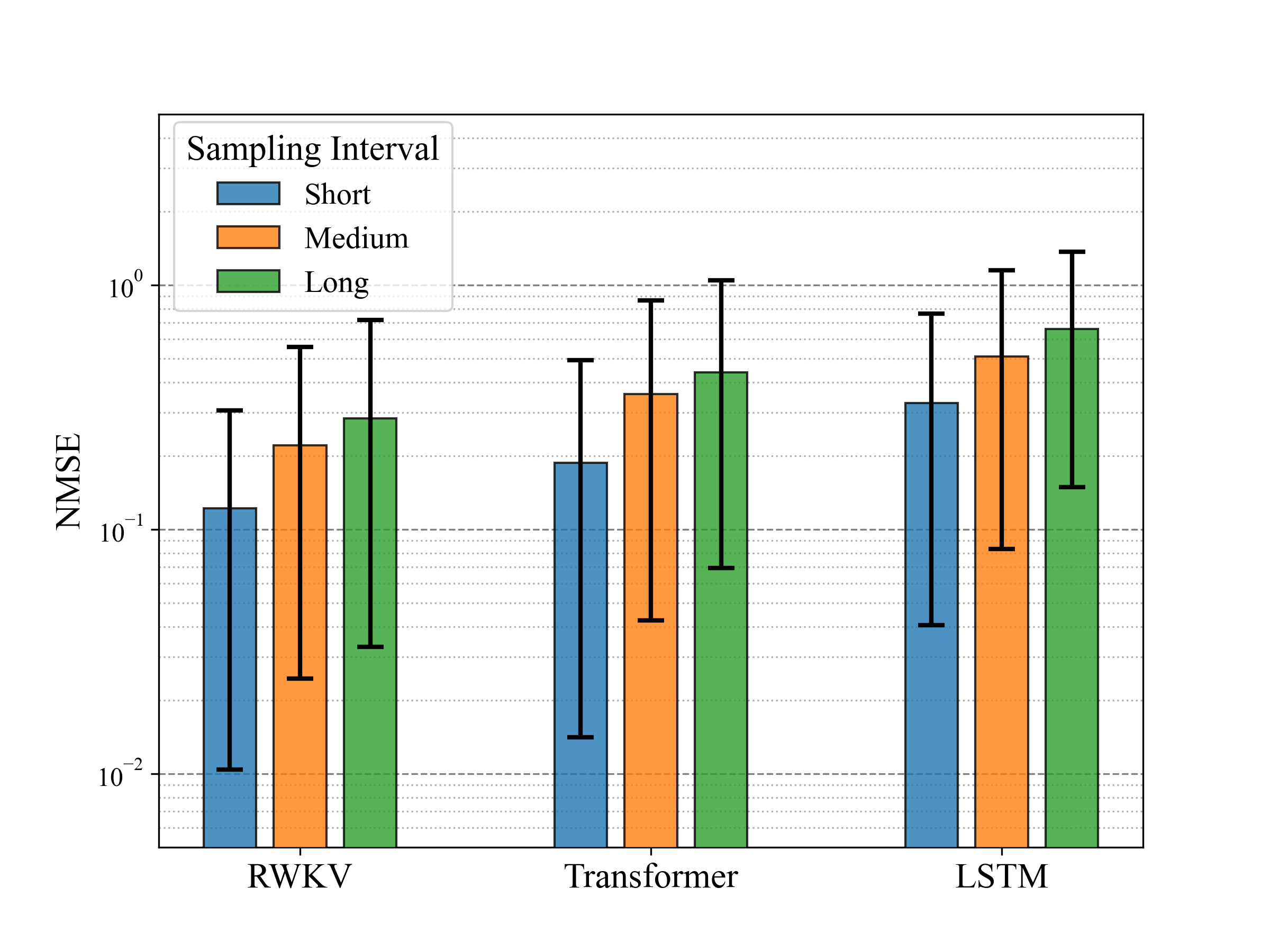} 
    \caption{NMSE under various sampling intervals.}
    \label{fig.4:interval}
\end{figure}

The results, presented in Fig.~\ref{fig.4:interval}, show that the NMSE for all models increases with the sampling interval. This performance degradation is attributed to the reduced data correlation that arises from the sparser sampling required under high mobility conditions. In this process, RWKV maintains the slowest error growth and the lowest error across all conditions. Even for long sampling intervals, it outperforms baselines by 1.88\textasciitilde3.66 dB gains. This stems from the deep interleaved architecture of RWKV, enabling it to fully exploit data dependencies and thereby exhibit the least sensitivity to interval changes. Moreover, RWKV consistently displays the shortest error bars, signifying its superior stability. 

\subsubsection{Robustness to Channel Noise}
\begin{figure}[!t]
    \centering 
    \includegraphics[width=0.45\textwidth, trim=0.28cm 0.18cm 0.5cm 1.8cm, clip=true]{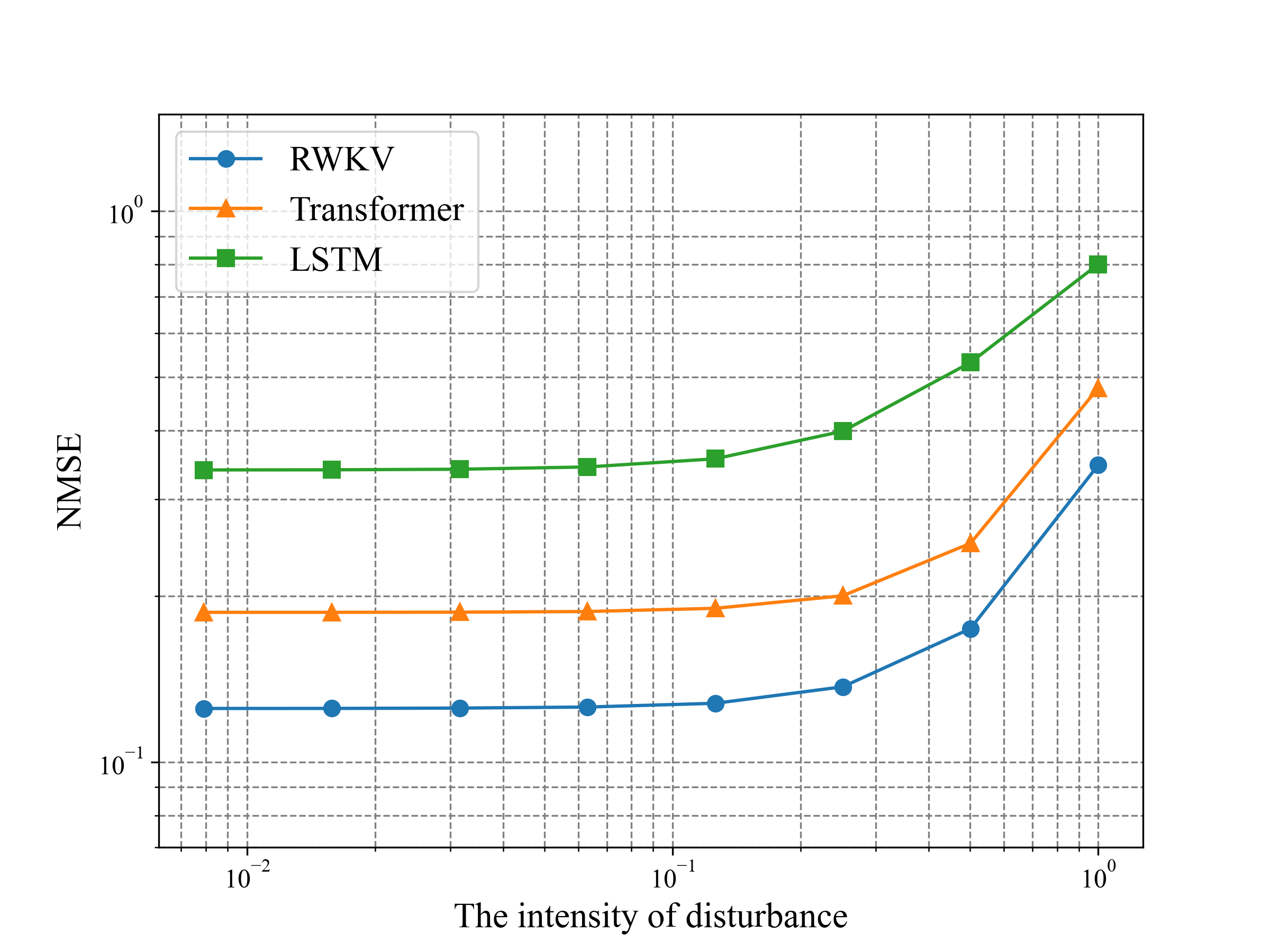} 
    \caption{Model robustness against input noise.}
    \label{fig.5:robustness}
\end{figure}

This part we evaluate the robustness of our proposed model from the perspective of external noise and internal error propagation. We first introduce noise of varying intensities into the test data to examine the model's resilience against external noise. As depicted in Fig.~\ref{fig.5:robustness}, the NMSE for all methods increases with noise intensity, with RWKV consistently maintaining the lowest error. This indicates the superior robustness of the proposed RWKV against noise, which can be explained by its precise and powerful modeling of global dependencies.

\begin{figure}[!t]
    \centering 
    \includegraphics[width=0.45\textwidth, trim=0.28cm 0.18cm 0.5cm 1.8cm, clip=true]{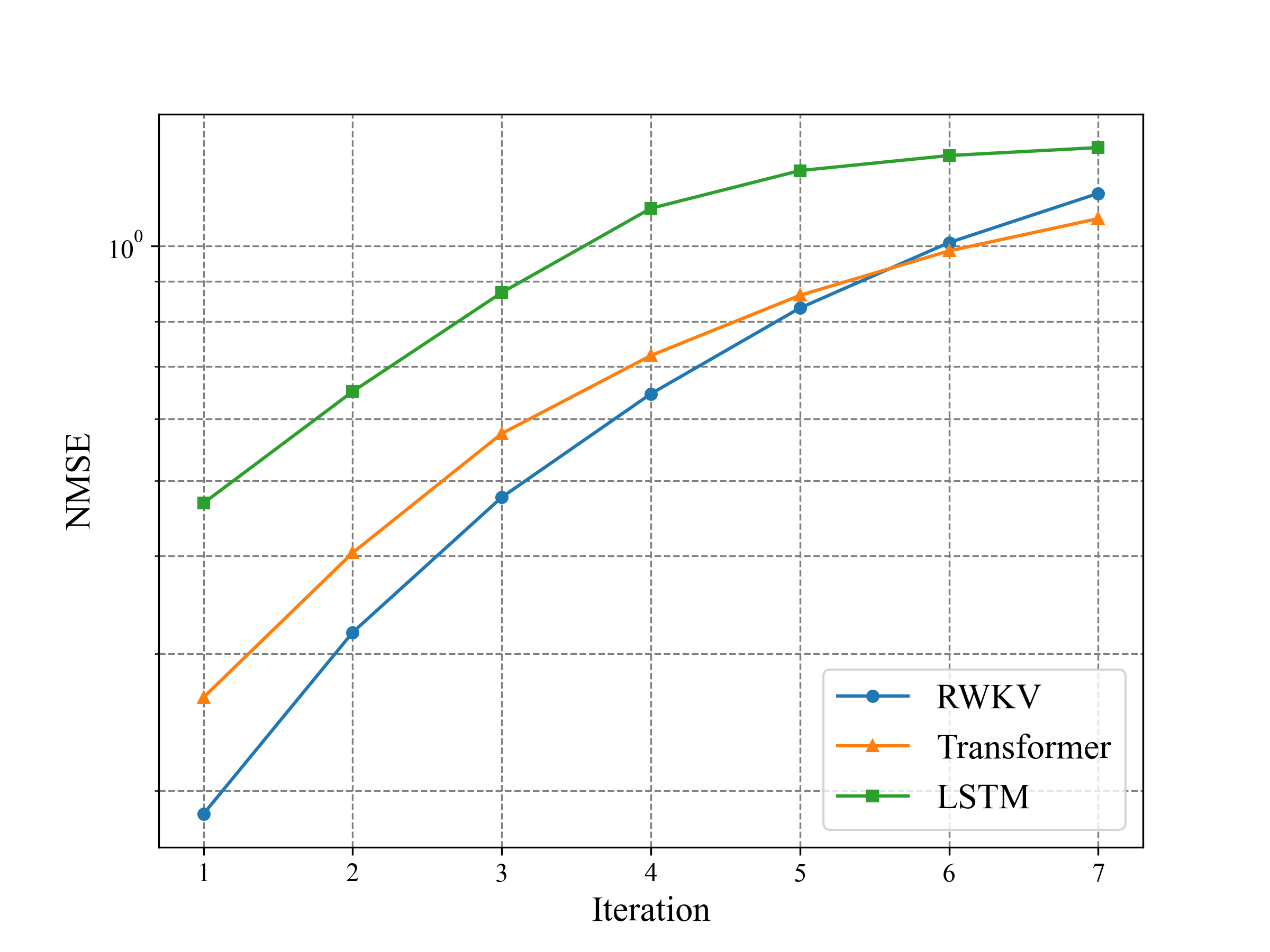}  
    \caption{Model robustness against error propagation in multi-iteration autoregressive test.}
    \label{fig.6:autoregressive}
\end{figure}

Then, we use a multi-iteration autoregressive test, which feeds the model's current prediction back as input for the subsequent time step. This test is intended to assess models’ ability to suppress cumulative error in long-term prediction. As illustrated in Fig.~\ref{fig.6:autoregressive}, RWKV consistently outperforms baselines. Both RWKV and Transformer predictions fail (NMSE > 1) at step 6, whereas LSTM fails earlier at step 4. Notably, within the effective prediction range, RWKV's performance advantage over Transformer diminishes progressively. We infer that while RWKV's time-decay mechanism provides a strong physics-intuitive inductive bias leading to superior initial performance, its heavy weighting of recent inputs simultaneously exacerbates cumulative error.

\section{Conclusion}\label{section5}
In this paper, we start from a task-driven perspective, introducing the emerging RWKV model to mobile channel prediction as it fulfills the task's requirements of global pattern learning and time order characterization, and performing necessary adaptations. Extensive experimental evaluations demonstrate the method's significant performance gains over existing methods, as well as its excellent robustness under challenging conditions such as high mobility and strong noise. These advantages stem from the RWKV model's deep synergy of global pattern learning and structural time-order characterization, which better aligns with the functional requirements of the channel prediction task. We hope that these findings and the task-driven model design concept will provide inspiration for more wireless AI tasks.

\section*{Acknowledgment}
This work was supported in part by National Natural Science Foundation of China under Grants 62394292 and 624B2129, 
in part by Zhejiang Provincial Key R\&D Program under Grant 2023C01021, 
and in part by the Fundamental Research Funds for the Central Universities No. 226-2024-00069. (\textit{Corresponding Author: Zhaoyang~Zhang})

\ifCLASSOPTIONcaptionsoff
  \newpage
\fi


\bibliographystyle{IEEEtran}
\bibliography{ref.bib}

\end{document}